\documentclass[twocolumn,aps,prl,floatfix,superscriptaddress,linenumbers]{revtex4}
\usepackage{tikz}
\usepackage{amsmath,amssymb}
\usepackage[colorinlistoftodos, color=green!40, prependcaption]{todonotes}
\usepackage{graphicx,float}
\usepackage{times,txfonts}
\usepackage{textcomp}
\usepackage{color}
\usepackage{multirow}
\usepackage{color}
\usepackage{soul}
\usepackage{hyperref}
\usepackage{natbib}
\usepackage{nicefrac}
\usepackage{multirow}
\usepackage{blindtext}
\usepackage[export]{adjustbox}
\usepackage{siunitx}
\usepackage[abbreviations]{glossaries-extra}
\usepackage{etoolbox}
\preto\section{\glsresetall}

\definecolor{hugoColor}{RGB}{59,134,255}

\newcommand{\fref}[1]{Fig.~\ref{#1}}

\DeclareRobustCommand{\autocircled}[1]{%
	\tikz[baseline=(char.base)]{
		\node[draw,circle,inner sep=0.5pt] (char) {#1};
	}%
}

\glssetcategoryattribute{abbreviation}{nohyper}{true}
\makeglossaries
\newabbreviation{eo}{EO}{electro-optic}
\newabbreviation{mzm}{MZM}{Mach-Zehnder modulator}
\newabbreviation{cpw}{CPW}{coplanar waveguide}
\newabbreviation{gsg}{GSG}{ground–signal–ground}
\newabbreviation{solt}{SOLT}{short–open–load–through}
\newabbreviation{imdd}{IMDD}{intensity-modulation and direct-detection}
\newabbreviation{dsp}{DSP}{digital signal processing}
\newabbreviation{snr}{SNR}{signal-to-noise ratio}
\newabbreviation{ltfs}{LT-on-FS}{lithium-tantalate-on-fused-silica}

\begin{document}
	
	\title{Low voltage and high-bandwidth thin-film lithium tantalate modulator on a silicon dioxide substrate}
	
	\author{Zihan Li}
	\thanks{Equal contribution for this work}
	\affiliation{Institute of Physics, Swiss Federal Institute of Technology, Lausanne (EPFL), CH-1015 Lausanne, Switzerland}
	\affiliation{Institute of Electrical and Micro Engineering (IEM), EPFL, CH-1015 Lausanne, Switzerland}
	
	\author{Alexander Kotz}
	\thanks{Equal contribution for this work}
	\affiliation{Institute of Photonics and Quantum Electronics (IPQ), Karlsruhe Institute of Technology (KIT), 76131 Karlsruhe, Germany}
	
	\author{Adrian Schwarzenberger}
	\affiliation{Institute of Photonics and Quantum Electronics (IPQ), Karlsruhe Institute of Technology (KIT), 76131 Karlsruhe, Germany}
	
	\author{Christian Koos}
	\affiliation{Institute of Photonics and Quantum Electronics (IPQ), Karlsruhe Institute of Technology (KIT), 76131 Karlsruhe, Germany}
	
	\author{Tobias J. Kippenberg}
	\email{tobias.kippenberg@epfl.ch}
	\affiliation{Institute of Physics, Swiss Federal Institute of Technology, Lausanne (EPFL), CH-1015 Lausanne, Switzerland}
	\affiliation{Institute of Electrical and Micro Engineering (IEM), EPFL, CH-1015 Lausanne, Switzerland}
	
	\begin{abstract}
	\end{abstract}
	
	\maketitle
	
	\begin{figure*}[t]
		\centering
		\includegraphics[width=\linewidth]{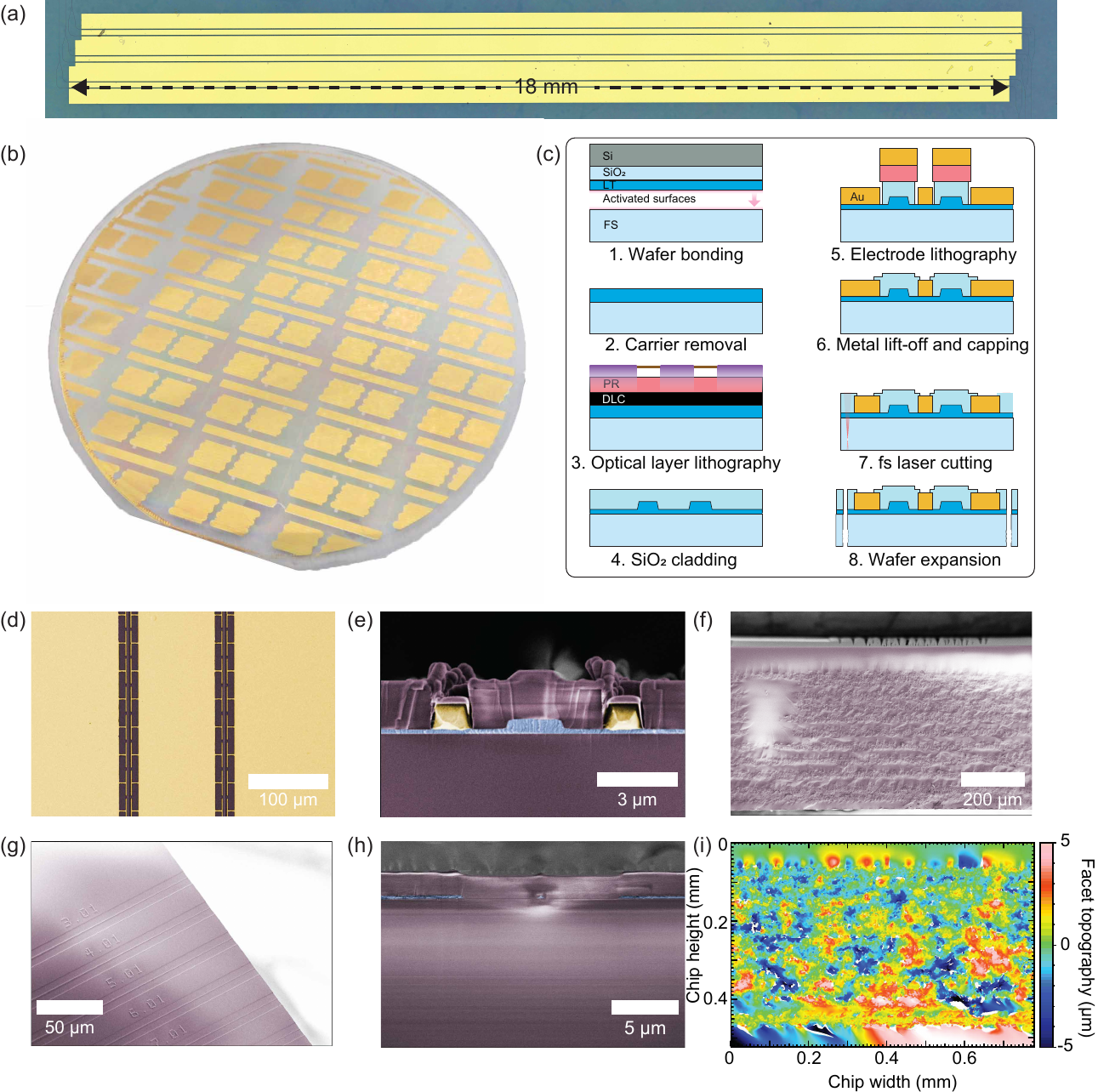}
		\caption{\textbf{Low voltage and high bandwidth thin film lithium tantalate modulators on a fused silica substrate.} {\bf (a)} Microscope picture of a chip comprising three modulators from a 4-inch wafer {\bf (b)}. {\bf (c)} Fabrication process flow for the fused-silica-based lithium tantalate modulator, including the substrate preparation, device fabrication, and chip singulation via a femtosecond laser. {\bf (d)-(e)} Colored scanning electron microscopy (SEM) image of a segmented coplanar waveguide and its cross section. {\bf (g)-(h)} SEM of the chip facet after laser singulation. The lithium tantalate rib waveguide is colored blue, gold electrodes are colored yellow, and the silicon dioxide is colored purple. {\bf (f)} and {\bf (i)} show the topography of the whole facet after the laser cutting by SEM and optical profilometer, respectively.}
		\label{fig:fig1}
	\end{figure*}

	\textbf{
		Modern communication networks demand ever-increasing transmission bandwidth, placing stringent requirements on low-cost, high-performance electro-optic modulators. Substantial advances have been made in integrated photonics employing lithium niobate on insulator. In contrast, photonic integrated circuits based on lithium tantalate---a material already commercially adopted for wireless filters---have been developed, offering reduced DC drift, higher optical power handling, and lower birefringence. These advantages enable more complex and dense photonic integrated circuits, and make lithium tantalate a promising material platform for next-generation integrated electro-optic modulators. However, in contrast to the extensively studied thin-film lithium niobate platform, thin-film lithium tantalate modulators have only been explored on silicon substrates.
		Here, we report the first fabrication and characterization of thin-film lithium tantalate electro-optic modulators manufactured on a 4-inch (100 mm) fused-silica substrate for adapting a low-loss slow-wave microwave electrode to improve the electro-optic bandwidth. By employing a slow-wave electrode design to achieve velocity matching between microwave and optical signals, the demonstrated modulator achieves a 3-dB electro-optic bandwidth of \SI{64}{\giga\hertz} with a low half-wave voltage of \SI{1.53}{\volt}, with potential to operate at the measured \SI{100}{\giga\hertz} electrical bandwidth, if the employed spectral biasing is removed. The modulator moreover exhibits low bias drift, with a constant switching voltage down to \SI{10}{\milli\hertz}. This performance enables high-speed data transmission comparable to state-of-the-art lithium niobate modulators fabricated on quartz substrates. Using the fabricated devices, a net single lane data rate of \SI{440.6}{\giga\bit\per\second} is achieved using PAM8 signaling. These results establish thin-film lithium tantalate as a viable and scalable alternative to lithium niobate for high-performance electro-optic links in next-generation communication systems.}

	\vspace{5mm}

	\section{Introduction}
	The rapid advancement and deployment of artificial intelligence have driven the demand for increasingly large-scale computing clusters. However, progress in interconnect technologies has lagged behind the growth of computational capacity, leading to a widening gap between processing throughput and network bandwidth~\cite{LightmatterOFC2025}. Conventional electrical interconnects are fundamentally limited in both transmission distance and bandwidth, making optical links indispensable for next-generation data-center and high-performance computing systems. Traditional optical modules, which rely on directly modulated semiconductor laser diodes to convert electrical signals into optical signals, are increasingly unable to meet the rising requirements for data rate and energy efficiency. As data centers scale up, power consumption associated with these modules has become a critical bottleneck.
	
	To address these challenges, broadband external optical modulators, such as silicon microring modulators and thin-film lithium niobate (TFLN) Mach–Zehnder modulators (MZMs), have been extensively investigated. By integrating with laser sources and digital signal processors, these modulators enable compact and high-speed pluggable optical modules for dense optical interconnects. Among various material platforms, lithium niobate ($\mathrm{LiNbO_3}$, LN), a well-established electro-optic crystal, is particularly attractive due to its broadband Pockels effect and low operating voltage, offering clear advantages over silicon photonic integrated circuits (PICs). Since the first demonstration of TFLN modulators~\cite{Wang:18}, they have achieved electro-optic (EO) \SI{3}{dB} bandwidths exceeding \SI{110}{\giga\hertz} with CMOS-compatible drive voltages\cite{Wang:18,Xu:20}. Nevertheless, their widespread adoption remains constrained by high fabrication costs and several practical challenges, including DC bias instability~\cite{Holzgrafe:24} and photorefractive effects under high optical power~\cite{xu2021mitigating}.
	
	Recently, lithium tantalate ($\mathrm{LiTaO_3}$, LT) photonic integrated circuits have been developed~\cite{WangNature:24} and high-speed modulators have been reported~\cite{WangOptica:24}, establishing LT as a promising alternative platform for electro-optical photonic integrated circuits. Already used in high volumes as RF filters, LT benefits from mature crystal growth techniques and an established foundry infrastructure, which are expected to support wafer-scale manufacturing at costs and volumes comparable to silicon-on-insulator platforms~\cite{soitec2021}. Compared with LN-based devices, LT-based electro-optic modulators exhibit improved DC stability and enhanced optical power handling, offering further advantages for long-haul and high-power transmission applications. In addition, LT exhibits a lower birefringence ($\delta n =\vert n_o-n_e\vert=0.004$) than LN ($\delta n = 0.074$~for LN at \SI{1550}{\nano\meter}, which simplifies the design of complex PICs, and enables arrayed waveguide gratings (AWG)~\cite{Hulyal:25}, which are challenging to implement in LN. Moreover, the low birefringence of LT enables wideband operation, as demonstrated for electro-optical frequency combs~\cite{zhang2025ultrabroadband}.
	On the lithium tantalate-on-insulator (LTOI) platform, \SI{3}{dB} EO modulation bandwidths exceeding \SI{110}{\giga\hertz} have been demonstrated on silicon substrates~\cite{WangOptica:24}. However, the high dielectric constant of silicon (11.7) fundamentally constrains the design of traveling-wave \gls{cpw} electrodes, particularly with respect to the capacitance for impedance and velocity matching. Therefore, in our previous work, on-chip microwave losses still limit the interaction length of LT-based modulators to approximately \SI{6}{\milli\meter}, resulting in a relatively high half-wave voltage ($V_\pi$) of around \SI{4.8}{\volt}~\cite{WangOptica:24}.
	To maintain efficient modulation, the electrode-to-waveguide gap must be kept within a few micrometers. Meanwhile, the impedance-matching condition determines the width of the \gls{cpw} signal conductor. These competing requirements result in increased conductor loss for the microwave signal, ultimately limiting the EO bandwidth. To overcome substrate-induced bandwidth limitations, Kharel \textit{et al.} and Chen \textit{et al.} proposed the use of low-permittivity substrates, such as quartz~\cite{Kharel:21} or released silicon~\cite{chen2022high}, combined with capacitively loaded slow-wave \gls{cpw} electrodes. The introduction of periodic T-shaped segments effectively separates the current paths in the \gls{cpw} and allows for a wider center conductor while preserving a \SI{50}{\ohm} impedance match without violating the velocity-matching condition. These design modifications significantly reduce microwave loss and increase the EO modulation bandwidth. Supplementary Note~2 explains how the microwave loss benefits from the wider signal conductor and the wider effective current separation. 
	
	Here, we further improve LT-based electro-optic modulators by replacing the silicon substrate with fused silica and implementing an optimized segmented electrode design. This approach enables a favorable balance between velocity matching, impedance matching, and microwave loss reduction. As a result, we achieve a \SI{64}{\giga\hertz} \SI{3}{dB} EO modulation bandwidth while maintaining a low $V_\pi$ of \SI{1.53}{\volt} in the C-band. The bandwidth is mainly limited by an inadvertent optical path imbalance at the modulator's output, which is used for spectral biasing. Optimized design projects a bandwidth expansion to \SI{100}{\giga\hertz} while preserving the same half‑wave voltage. To demonstrate the practical application of the proposed devices, we perform high-speed \gls{imdd} signaling experiments, achieving a net data rate of \SI{440.6}{\giga\bit\per\second}, which is comparable to values reported for state-of-the-art lithium niobate modulators \cite{Berikaa_TFLN}.

	\section{Results}
	
	\subsection{Design and fabrication}
	The LT on fused silica (LT-on-FS) electro-optic modulators are fabricated from a commercial X-cut thin-film lithium tantalate-on-insulator (LTOI) wafer (NANOLN), consisting of a \SI{600}{\nano\meter}-thick $\mathrm{LiTaO_3}$ single crystal thin-film, a \SI{2}{\micro\meter}-thick silicon dioxide buffer layer, and a \SI{525}{\micro\meter}-thick silicon handle substrate. To enable low-loss microwave propagation, the $\mathrm{LiTaO_3}$ thin-film is transferred to a 4-inch, \SI{500}{\micro\meter}-thick fused silica wafer via direct wafer bonding, with oxygen plasma activation employed as the surface pretreatment (EVG 810, EVG 501, and EVG 301). The original silicon carrier and the silicon dioxide buffer layer are subsequently removed through a combination of backside grinding, fluorine-based plasma dry etching, and buffered hydrofluoric acid wet etching.
	
	With an optimized bonding recipe and carrier removal process, the fabricated LT-on-FS wafer achieves a film transfer yield exceeding $\sim$95\%. In addition, the relatively small thermal expansion coefficient mismatch between silicon ($2.6\times10^{-6}$~K$^{-1}$) and fused silica ($0.5\times10^{-6}$~K$^{-1}$) allows higher annealing temperatures during the bonding process compared with direct thin-film transfer from ion-implanted bulk LT wafers, which exhibit pronounced large and anisotropic thermal expansion ($\alpha_z = 2.0\times10^{-6}$~K$^{-1}$, $\alpha_{x,y} = 15.0\times10^{-6}$~K$^{-1}$\cite{kim1969thermal}).
	
	Following the wafer preparation, the $\mathrm{LiTaO_3}$ photonic integrated circuits (PICs) are fabricated using a direct etching approach previously reported in~\cite{Li:23, WangNature:24}. Deep-ultraviolet (DUV) stepper lithography (ASML PAS 5500/350C) is employed to define the photonic patterns, which are transferred into a diamond-like carbon (DLC) hard mask via oxygen plasma dry etching (SPTS APS). Then, the patterns are etched into the $\mathrm{LiTaO_3}$ layer using argon ion beam etching (Veeco Nexus IBE350). To remove redeposited $\mathrm{LiTaO_3}$ residues resulting from ion milling, a subsequent wet etching using an aqueous solution of hydrogen peroxide and potassium hydroxide is performed. The final etch depth is \SI{440}{\nano\meter}, leaving a \SI{160}{\nano\meter}-thick slab to ensure efficient electro-optic modulation. A second DUV lithography and ion beam etching step is used to define the double-layer taper to reduce the edge coupling loss. The PICs are clad with a \SI{1.5}{\micro\meter}-thick $\mathrm{SiO_2}$ layer deposited by hydrogen-free high-density plasma-enhanced chemical vapor deposition (HD-PECVD)\cite{Qiu:2024}, which also serves as a sacrificial layer for metalization process. The microwave electrodes are patterned by DUV stepper lithography and fabricated through a sequence of dielectric etching, metal evaporation (\SI{15}{\nano\meter} Ti / \SI{800}{\nano\meter} Au), and lift-off \cite{WangNature:24}. To protect the soft gold electrodes, a \SI{200}{\nano\meter}-thick $\mathrm{SiO_2}$ cap layer is deposited by PECVD, and then the pad area is opened by wet etching. Finally, the wafer is singularized using femtosecond-laser stealth dicing (General Intelligent Equipment Co., Ltd.). The key fabrication steps are summarized in \fref{fig:fig1}~(c).
	
	Figure~\ref{fig:fig1}~(a) shows a fabricated \SI{3}{\milli\meter}$\times$\SI{20}{\milli\meter} chip containing three unbalanced \gls{mzm}, each composed of a pair of \SI{18}{\milli\meter}-long modulation arms and two 50:50 multimode interference (MMI) couplers. The chip is taken from a 4-inch wafer, as shown in \fref{fig:fig1}~(b). Colored scanning electron microscopy (SEM) images reveal well-defined segmented \gls{cpw} and LT waveguides in both top-view (\fref{fig:fig1}~(d)) and cross-sectional (\fref{fig:fig1}~(e)) perspectives.
	
	Due to the amorphous structure of fused silica and the lack of an efficient deep reactive ion etching process, die singulation is performed using stealth dicing, in which arrays of modified points are generated inside the fused silica substrate by a femtosecond laser. These modified regions allow controlled chip separation during wafer expansion. Figures~\ref{fig:fig1}(f)–(h) show the chip facets, which are clean and intact. The smooth region near the top functional layer is suitable for direct fiber packaging without additional polishing, resulting in a fiber-to-fiber coupling loss of approximately \SI{12}{\decibel}. The facet topography is quantitatively characterized using an optical profilometer (Sensofar S-Neox), presenting a surface roughness of $\pm$\SI{5}{\micro\meter} induced by laser modification, which indicates the need for facet polishing before fiber arrays or laser diode coupling.
	
	\begin{figure*}[t]
		\centering
		\includegraphics[width=\linewidth]{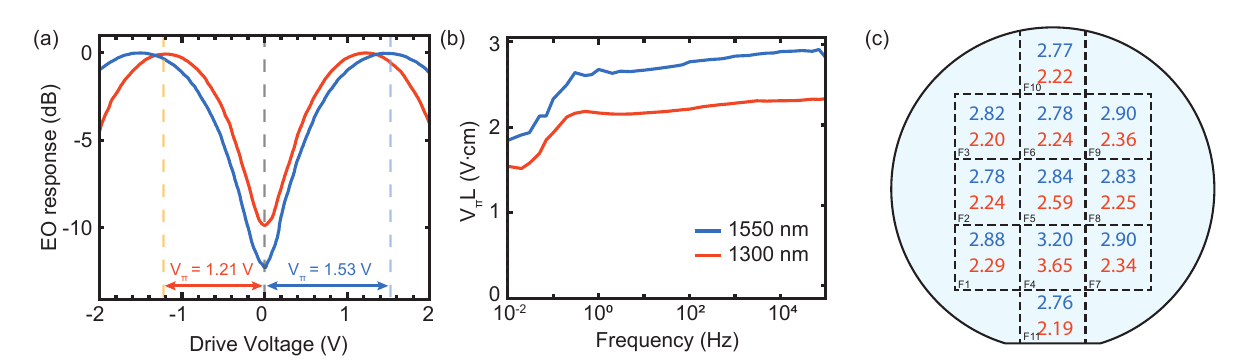}
		\caption{\textbf{Modulation efficiency characterization of LT-on-FS Mach-Zehnder modulators in the C- and O-band.} Measurements at wavelength of \SI{1550}{\nano\meter} are colored blue, and those at \SI{1300}{\nano\meter} are shown in orange. {\bf (a)}~Normalized optical transmission in logarithmic scale of an \SI{18}{\milli\meter}-long modulator as a function of the drive voltage at \SI{100}{\hertz}. The half-wave voltages $V_\pi$ are \SI{1.53}{\volt} and \SI{1.21}{\volt} at carrier wavelengths of \SI{1550}{\nano\meter} and \SI{1300}{\nano\meter}, respectively. {\bf (b)} Half-wave voltage-length $V_\pi L$ products as a function of drive voltage frequency. The electro-optic response is stable at both carrier wavelengths \SI{1550}{\nano\meter} and \SI{1300}{\nano\meter} for modulation frequencies from \SI{1}{\hertz} to \SI{10}{\kilo\hertz}. {\bf (c)}~Distribution of the measured $V_\pi L$ for the same device design across a 4-inch wafer.}
		\label{fig:fig2}
	\end{figure*}

	
	To achieve broadband electro-optic modulation, segmented T-shaped \gls{cpw} are designed. The segment length $L_\mathrm{S}$ (\fref{fig:fig3}~(a)) is varied to tune the microwave phase velocity, where increasing the segment length effectively reduces the microwave phase velocity. The bandwidth is maximized when the microwave phase velocity matches the optical group velocity, while simultaneously the characteristic impedance is matched to external circuitry~\cite{Kharel:21}. The electrode parameters, as defined in \fref{fig:fig3}~(a), selected for the presented device are ($W_\mathrm{T},~L_\mathrm{T},~W_\mathrm{S},~L_\mathrm{S},~G,~W_\mathrm{sig},~P$) = (1.5, 32, 2, 5, 5, 100, 35)~\SI{}{\micro\meter}.

	\subsection{Electro-optic Modulation}
	To evaluate the wafer-scale modulation performance of the fabricated devices, we select modulators with identical designs from each exposure field and characterize their modulation efficiency, as the half-wave voltage–length product ($V_\pi L$). For low-frequency characterization, a sawtooth voltage waveform with a peak-to-peak amplitude ($V_\mathrm{pp}$) of \SI{10}{\volt} is applied to the modulator electrodes, while the corresponding optical output is simultaneously recorded using a photodetector and an oscilloscope. The half-wave voltage is calculated by fitting the EO response curve.
	
	Owing to the small difference in the optical group index between the O-band (e.g. \SI{1300}{\nano\meter}) and the C-band (e.g. \SI{1550}{\nano\meter}), the modulators are inherently compatible with both wavelength ranges, with minor sacrifice in insertion loss and extinction ratio (ER). To quantify the dual-band modulation capability, we measure the modulation efficiency using laser sources at \SI{1300}{\nano\meter} and \SI{1550}{\nano\meter} separately. Figure~\ref{fig:fig2} summarizes the dual-wavelength measurement results, with data acquired at \SI{1300}{\nano\meter} shown in orange and at \SI{1550}{\nano\meter} shown in blue. Figure~\ref{fig:fig2}~(a) presents the normalized transmission over one modulation period for a representative device (C1\_F11\_1.02) on a logarithmic scale. For an \SI{18}{\milli\meter}-long modulator, half-wave voltages of \SI{1.53}{\volt} and \SI{1.21}{\volt} are extracted at wavelengths of \SI{1550}{\nano\meter} and \SI{1300}{\nano\meter}, respectively. The corresponding extinction ratios are approximately \SI{12}{\decibel} in the C-band and \SI{10}{\decibel} in the O-band.
	
	
	\begin{figure*}[t]
		\centering
		\includegraphics[width=\linewidth]{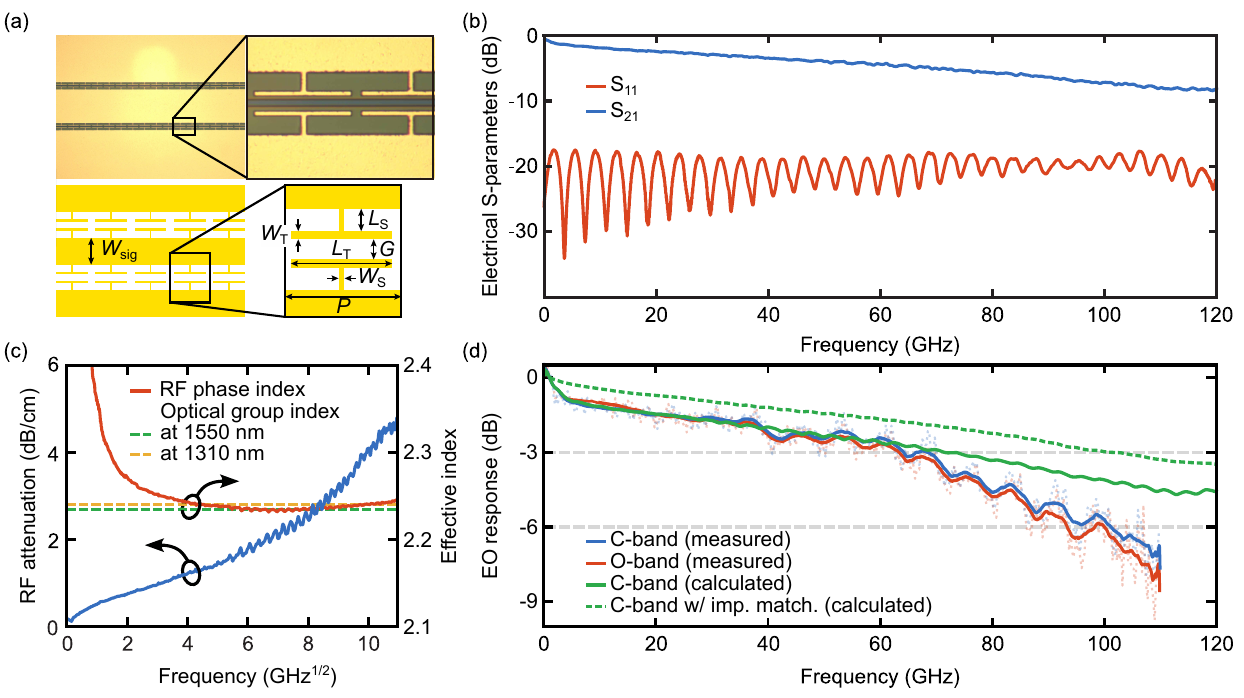}
		\caption{
			\textbf{Characterization of the modulator frequency response.} {\bf(a)} Optical micrograph of the slow-wave electrodes of the \glsentryfull{mzm} with an inset showing the T-shaped capacitive loading elements and with a schematic illustrating the design parameters. {\bf (b)} Measured electrical S-parameters of the \SI{18}{\milli\meter}-long capacitively loaded \glsentryfull{cpw} electrodes of the \gls{mzm}. {\bf (c)} Extracted microwave attenuation of the \gls{cpw} (blue) and effective phase index (red) of the RF signal plotted as a function of the square root of the frequency. As a reference, the effective group refractive index of the optical signal at a wavelength of \SI{1550}{\nano\meter} (green) and \SI{1310}{\nano\meter} (yellow) are shown. {\bf (d)} Frequency-dependent electro-optic (EO) response of the \SI{18}{\milli\meter}-long \gls{mzm} for carrier wavelengths of \SI{1550}{\nano\meter} (C-band, blue) and \SI{1310}{\nano\meter} (O-band, red). The light-colored dots correspond to raw data, while the solid-colored lines indicate smoothed data obtained by applying a centered \SI{5}{\giga\hertz}-wide moving-average filter. The data are obtained from vector network analyzer (VNA) measurements with a high-speed photodiode, de-embedding, and normalization to the response at \SI{1}{\giga\hertz}. For the O-band measurement, the data points below \SI{10}{\giga\hertz} were obtained with a separate low-speed photodiode using the method described in \cite{Nagarajan_ACVpi1}. We used this approach to avoid uncertainties of the transfer characteristics of the high-speed photodiode when operated at low modulation frequencies in the O-band. The solid green curve indicates the EO response of our \gls{mzm} calculated from the extracted transmission line parameters and the simulated effective group refractive index of the optical wave in the C-band, see Subfigure~(c) according to~\cite{lin_high-throughput_1987}. The deviation from the measurement above \SI{60}{\giga\hertz} originates from an inadvertent imbalance in the optical paths before the combiner, which acts as an optical filter reducing the EO response. The dashed green curve shows the predicted performance of our \gls{mzm} with an optimized design, where the line impedance is perfectly matched to the source and the termination, and where the EO response is limited only by the microwave losses of the electrodes.
		}
		\label{fig:fig3}
	\end{figure*}

	We further analyze the dynamic behavior of our \gls{ltfs} \gls{mzm}. The device is designed as a traveling-wave electro-optic modulator such that the frequency response is primarily dominated by the RF loss of the transmission line, the matching between the group velocity of the optical wave and the phase velocity of the RF wave, as well as the impedance matching of the transmission line to the source and the termination~\cite{lin_high-throughput_1987}. The electrical characterization of the \gls{ltfs} \gls{mzm} is performed using a high-speed vector network analyzer (VNA; ME7838AX, Anritsu Corporation) covering a frequency range from \SI{70}{\kilo\hertz} to \SI{125}{\giga\hertz}. The measurements are carried out using a pair of \SI{110}{\giga\hertz} ground–signal–ground (GSG) RF probes (T110A-GSG0100, MPI Corporation). Prior to the measurement, a two-port short–open–load–through (SOLT) calibration is conducted on a commercial calibration substrate (AC2-2, MPI Corporation), setting the reference planes at the probe tips with a characteristic impedance of \SI{50}{\ohm}. The electrical scattering parameters of the \SI{18}{\milli\meter}-long MZM are then obtained by probing the input and output contact pads.
	
	As shown in \fref{fig:fig3}~(b), the segmented \gls{cpw} exhibits a reflection coefficient ($S_{11}$) below \SI{-18}{\decibel}, and the electrical transmission coefficient ($S_{21}$) drops to \SI{-8.2}{\decibel} at \SI{120}{\giga\hertz}. This corresponds to a microwave loss of approximately \SI{4.6}{\decibel\per\centi\meter} at \SI{120}{\giga\hertz}, highlighting the low RF attenuation of the T-shape segmented electrodes, see blue curve in \fref{fig:fig3}~(c). We also extract the frequency-dependent effective phase refractive index of the RF signal from the S-parameters of the \gls{cpw}, see red curve in \fref{fig:fig3}~(c). The strong increase at low frequencies is attributed to the finite conductivity of the electrodes, which allows magnetic fields to penetrate the conductors and thereby increases the internal inductance~\cite{Ke_low_frequency_1995}. For comparison, the effective group refractive indices for the optical wave, obtained from numerical simulations, are given at a wavelength of \SI{1550}{\nano\meter} (C-band, green) and \SI{1310}{\nano\meter} (O-band, yellow). The results indicate good velocity matching in both bands.
	
	To measure the electro-optic frequency response of the LT-on-FS MZM, an optical carrier is coupled to the modulator, and the device is driven by the first VNA port, while the second port is connected to a photodiode that detects the modulated optical signal. To avoid reflections of the RF wave, the \gls{mzm} is terminated by a second probe and a \SI{50}{\ohm} resistor. To isolate the electro-optic frequency response of the \gls{ltfs} \gls{mzm}, we use appropriate de-embedding techniques that shift the reference planes of the VNA measurement to the tips of the input probe and the optical output interface of the modulator. These de-embedding steps rely on the frequency responses of the second RF probe and the photodiode as provided by the manufacturers.
	
	\fref{fig:fig3}~(d) shows the electro-optic frequency response of the \SI{18}{\milli\meter}-long \gls{ltfs} \gls{mzm} normalized to its value at \SI{1}{\giga\hertz} for optical carrier wavelengths of \SI{1550}{\nano\meter} (C-band, blue) and \SI{1310}{\nano\meter} (O-band, red). The light-colored dots represent the raw measurement data, while the solid-colored lines correspond to the same data after applying a centered \SI{5}{\giga\hertz}-wide moving-average filter. The smoothed device characteristics in the C-band indicate a \SI{3}{dB} bandwidth of \SI{64}{GHz} and a \SI{6}{dB} bandwidth that exceeds \SI{100}{GHz}. The measurements in the O-band show similar characteristics, with some uncertainties originating from the O-band calibration of the photodiode. The increased roll-off of the electro-optic response above \SI{60}{\giga\hertz} is attributed to an inadvertent imbalance in the optical paths after the modulation and before the combiner, which introduces an additional optical filtering effect. This can be easily avoided by an adapted design.
	
	The solid green curve in \fref{fig:fig3}~(d) shows the electro-optic response of the \gls{ltfs} \gls{mzm} calculated from the extracted transmission-line parameters and the simulated effective group refractive index of the optical wave in the C-band (see \fref{fig:fig3}~(c)) according to~\cite{lin_high-throughput_1987}. The prediction agrees with the measured electro-optic response of our \gls{mzm} at modulation frequencies below \SI{60}{\giga\hertz} and provides a reference for the response for a modulator without the imbalance. Further analysis of the initial drop of the electro-optic response at low frequencies reveals a mismatch of the characteristic impedance of the modulator, which amounts to approximately \SI{42}{\ohm} according to the electric S-parameters, to the \SI{50}{\ohm} source and termination impedance. An \SI{18}{mm}-long \gls{mzm} with optimized design, for which the electro-optic response is limited only by the microwave losses of the electrodes, could achieve a \SI{3}{dB} bandwidth of \SI{100}{\giga\hertz} on our \gls{ltfs} platform, as indicated by the dashed green curve in \fref{fig:fig3}~(d)).
	

	\subsection{High-Speed IMDD Signaling with LT-on-FS MZM}
	
	\begin{figure*}[t]
		\centering
		\includegraphics[width=\linewidth]{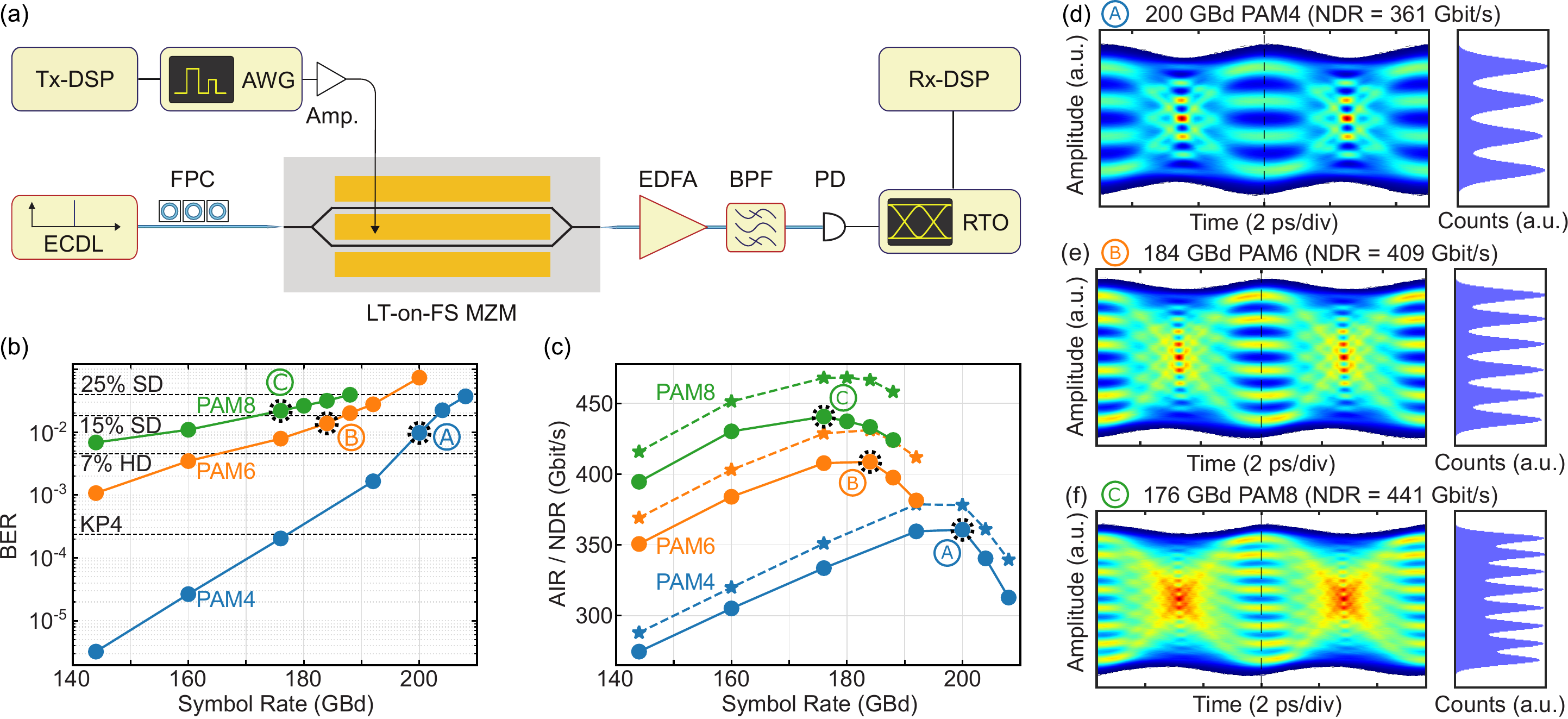}
		\caption{\textbf{\Glsentryfull{imdd} signaling experiment using our \glsentryfull{ltfs} \glsentryfull{mzm}.} \textbf{(a)} Experimental setup: An external cavity diode laser (ECDL) provides the optical carrier, and a fiber polarization controller (FPC) is used to adjust the polarization. Optical coupling at the input and output of the LT-on-FS chip is achieved via a pair of lensed fibers. The electrical drive signals are synthesized by transmitter digital signal processing (Tx-DSP), generated by an arbitrary waveform generator (AWG), and amplified by a broadband RF amplifier (Amp.). To enable reception by a rather insensitive photodiode (PD) without electrical amplifiers, the modulated optical signal is boosted by an erbium-doped fiber amplifier (EDFA), and out-of-band amplified spontaneous emission (ASE) noise is suppressed by a tunable bandpass filter (BPF). The output signal of the PD is digitized by a high-speed real-time oscilloscope (RTO) and processed offline using the receiver DSP (Rx-DSP). \textbf{(b)} Measured bit error ratios (BER) as a function of the symbol rate for pulse-amplitude modulated signals with eight (PAM8, green), six (PAM6, orange), and four (PAM4, blue) levels. The horizontal black dashed lines indicate the thresholds for \SI{25}{\percent} and \SI{15}{\percent} soft-decision forward error correction (SD-FEC), as well as \SI{7}{\percent} and \SI{5.8}{\percent} (KP4) hard-decision forward error correction (HD-FEC). The labels \autocircled{A}, \autocircled{B}, and \autocircled{C} refer to the corresponding eye diagrams in Subfigures (d) - (f). \textbf{(c)}~Achievable information rates (AIR, dashed lines) and corresponding net data rates (NDR, solid lines) vs. symbol rate. A maximum NDR of \SI{441}{\giga\bit\per\second} is achieved for PAM8 signals at a symbol rate of \SI{176}{GBd}. \textbf{(d)~-~(f)}~Eye diagrams after Rx-DSP and corresponding histograms, taken at the center of the respective symbol slot (indicated by the vertical dashed line), for symbol rates offering the highest NDR for each modulation format. These data points are highlighted in Subfigures~\textbf{(b)} and \textbf{(c)}).}	
		\label{fig:fig4}
	\end{figure*}
	
	To demonstrate the transmission performance of our \gls{ltfs} \gls{mzm}, we perform a high-speed \gls{imdd} signaling experiment with the setup illustrated in \fref{fig:fig4}~(a). An external-cavity diode laser (ECDL) provides the optical carrier at a wavelength of \SI{1550}{\nano\meter} with an output power of \SI{17.8}{dBm}. The polarization of the light is adjusted via a fiber-based polarization controller (FPC) to excite the quasi-transverse electric mode of the LT waveguide, having the dominant component of the electric field parallel to the substrate plane. Lensed fibers are used for optical coupling to and from the chip. The electrical drive signal is generated using a high-speed arbitrary waveform generator (AWG; M8199B, Keysight Technologies Inc.) with a sampling rate of \SI{256}{GSa\per\second} and transmitted via a \SI{20}{\centi\meter}-long RF cable, a broadband RF amplifier (Amp.; AH15199B, Anritsu Corporation) and a \SI{110}{\giga\hertz} RF probe to the \gls{cpw} of the \gls{mzm}. Pulse-amplitude modulated (PAM) signals with various levels are generated from pseudo-random bit sequences (PRBS) with a pattern length longer than $2^{17}$~bits using offline digital signal processing (Tx-DSP). The Tx-DSP chain furthermore consists of a root-raised cosine pulse shaping filter with a roll-off factor of $\beta = 0.05$, predistortion to account for the frequency response of the transmitter electronics excluding the modulator, and clipping of the signal to reduce the peak-to-average power ratio to a value of \SI{9}{dB}. The predistortion is based on linear minimum-mean-square-error equalization. At the output, the \gls{cpw} is terminated with a \SI{50}{\ohm} resistor via a second \SI{110}{\giga\hertz} RF probe and a broadband bias-tee (not shown) to set the quadrature operation point of the \gls{mzm} for intensity modulation. The modulated optical signal has a power of \SI{3.7}{dBm} in the fiber right after the \gls{mzm}, exceeding typical specifications for high-speed optical Ethernet transceivers~\cite{IEEE_ethernet}. Still, we had to use an erbium-doped fiber amplifier (EDFA) to slightly boost the optical output power to a level of approximately \SI{7.7}{dBm} at the receiver to allow for detection with a high-speed photodiode (PD; Fraunhofer HHI) and a directly attached real-time oscilloscope (RTO; UXR 1004A, Keysight Technologies Inc.). The latter has a sampling rate of \SI{256}{GSa\per\second} and a bandwidth of \SI{104}{\giga\hertz}. Note that, in practical systems, a broadband RF amplifier after the photodiode could be used, thus eliminating the need for the EDFA. The out-of-band amplified spontaneous-emission (ASE) noise of the EDFA is suppressed by a tunable bandpass filter (BPF). After digitization by the RTO, the data is finally extracted using offline DSP (Rx-DSP), which contains resampling to two samples per symbol, timing recovery, linear Sato equalization, and additional least-mean-square (LMS) equalization.
	
	In our signaling experiments, we generate and receive PAM signals with four (PAM4), six (PAM6) and eight (PAM8) power levels at symbol rates between \SI{144}{GBd} and \SI{208}{GBd}. Figure~\ref{fig:fig4}~(b) shows the bit error ratio (BER) as a function of the symbol rate, with dashed lines indicating thresholds for soft-decision forward-error correction (SD-FEC) with \SI{25}{\percent} and \SI{15}{\percent} overhead~\cite{GraelliAmat2020}, as well as for hard-decision forward-error correction (HD-FEC) with \SI{7}{\percent}~\cite{itu_g.975.1} and \SI{5.8}{\percent} (KP4)~\cite{IEEE_ethernet} overhead. The measured BER remain below the \SI{25}{\percent} SD-FEC limit for PAM8 signals at \SI{184}{GBd}, PAM6 signals at \SI{192}{GBd}, and PAM4 signals at \SI{208}{GBd}. Up to a symbol rate of \SI{176}{GBd}, the BER for PAM4 signals stays below the KP4 limit.
	
	To quantify the achievable information rate (AIR), the generalized mutual information (GMI) of our measurements is calculated using log-likelihood ratios between the transmitted and received signals while assuming a linear channel with additive white Gaussian noise (AWGN) as the only impairment~\cite{gmi_ivanov}. The results are indicated as dashed lines in \fref{fig:fig4}~(c), where the highest AIR of \SI{468.1}{Gbit\per\second} is achieved for PAM8 signals at a symbol rate of \SI{180}{GBd}. To estimate practically achievable net data rates (NDRs), penalties introduced by typical FEC codes must also be considered. To this end, the GMI is normalized by the number of bits that can be encoded into a single symbol and suitable FEC codes with the nearest lower normalized GMI threshold as given in \cite[Table 2]{Hu2022fec} are selected. The NDR is computed as the product of line rate and FEC code rate and is indicated by circles and solid lines in \fref{fig:fig4}~(c). For each modulation format, we identify the symbol rates at which the highest NDR are achieved and extract the corresponding eye diagrams as obtained after RX-DSP, see \fref{fig:fig4}~(d-f). The plots also show the corresponding histograms, evaluated at the center of the symbol period. The corresponding data points are highlighted in \fref{fig:fig4}~(b) and \fref{fig:fig4}~(c). Clear eye openings and distinct clustering at the different signal levels are observed. The highest NDR of \SI{440.6}{Gbit\per\second} is obtained for PAM8 signals at a symbol rate of \SI{176}{GBd} using SD-FEC with \SI{18.7}{\percent} overhead. For PAM4 signals at a symbol rate of \SI{176}{GBd}, the NDR is \SI{332.6}{Gbit\per\second} using KP4 FEC. These results are comparable to those reported for high-bandwidth LNOI \glspl{mzm}~\cite{Berikaa_TFLN}. Note that, at high symbol rates, the quality of the generated optical signals is primarily limited by the electrical signal source and not so much by the \gls{ltfs}, see Supplementary Note~1 for details.
	
	\section{Conclusion}
	Lithium niobate modulators have represented the state of the art in terms of electro-optic bandwidth and modulation efficiency. Whether lithium tantalate, an emerging electro-optic photonic integrated circuits platform, can serve as a potential replacement for lithium niobate has remained an open question due to concerns regarding microwave loss and its relatively higher dielectric constant. In this work, we adopt a design similar to that of advanced lithium niobate modulators and, for the first time, demonstrate thin-film lithium tantalate modulators fabricated on a fused silica substrate, which achieve performance comparable to lithium niobate counterparts.
	
	By transferring X-cut lithium tantalate (LT) thin films onto a fused silica (FS) wafer and implementing optimized segmented traveling-wave electrodes, we simultaneously achieve low microwave loss, precise velocity matching, and impedance matching. The fabricated LT-on-FS modulators exhibit uniform performance across a 4-inch wafer, with an average modulation efficiency of 2.86 (2.76, 3.20) \si{\volt\centi\meter} in the C-band and 2.42 (2.19, 3.65)~\si{\volt\centi\meter} in the O-band. While the measured \SI{3}{\decibel} electro-optic bandwidth of an \SI{18}{\milli\meter}-long Mach–Zehnder modulator is only \SI{64}{\giga\hertz}, the fundamental bandwidth limitation due to microwave losses of the traveling-wave electrodes indicates an achievable bandwidth of \SI{100}{\giga\hertz} with \gls{mzm} of the same length on the \gls{ltfs} platform. The observed roll-off in the electro-optic response is primarily due to an inadvertent imbalance in the optical path at the \gls{mzm} output. Beyond device-level characterization, we validate the system-level application of the LT-on-FS platform through high-speed \gls{imdd} experiments. Using PAM8 signaling in the C-band, a net data rate of up to \SI{440.6}{\giga\bit\per\second} is achieved, comparable to the performance reported for state-of-the-art thin-film lithium niobate modulators.
	
	Our results demonstrate that LT-on-FS modulators constitute a scalable, high-performance alternative to lithium niobate for broadband and energy-efficient electro-optic links. Our work provides a strong reference for the adoption of lithium tantalate in next-generation high-speed optical interconnects and its potential to support cost-effective, wafer-scale manufacturing for future communication systems.

	\vspace{1 EM}
	
	\noindent {\textbf{Data and Code Availability}} {Data and code used to produce the plots within this paper will be available at \texttt{Zenodo} upon publication of the manuscript.}
	
	\vspace{1 EM}
	
	
	
	\noindent \textbf{Funding}
	
	This work was supported by the Horizon Europe EIC Transition programme under grant agreement No. grant agreement 101131069 (ELLIPTIC) and under grant No. 101113260 (HDLN), as well as funding from the Swiss State Secretariat for Education, Research and Innovation (SERI). This work has received funding from the European Research Council (ERC) under the Horizon Europe research and innovation programme, grant agreement No. 101167540 (ATHENS), as well as from the German Research Foundation via the projects PACE (No. 403188360) and GOSPEL (No. 403187440).
	
	\vspace{1 EM}
	\noindent \textbf{Acknowledgements}
	We acknowledge the EPFL Center of MicroNano Technology (CMi) and the Institute of Physics (IPHYS) cleanroom for supporting on sample fabrication.
	
	\vspace{1 EM}
	
	\noindent\textbf{Competing Interests}
	C.K. and T.J.K. are co-founders and shareholders of Luxtelligence SA, St. Sulpice, Switzerland, a company engaged in electro-optic modulators based on ferroelectric materials, such as lithium niobate and lithium tantalate. The other authors declare no competing interests.

	\vspace{5mm}
	
	\bibliography{FSLTOIBib.bib}
	
\end{document}